\documentclass[12pt]{article}
\usepackage{latexsym}
\usepackage{epsfig}

\def\bg#1{\mbox{\boldmath$#1$}}

\newcommand{\del}{\partial}
\newcommand{\beq}{\begin{eqnarray}}
\newcommand{\eeq}{\end{eqnarray}}
\newcommand{\be}{\begin{eqnarray*}}
\newcommand{\ee}{\end{eqnarray*}}
\newcommand{\bk}{{\bf k}}
\newcommand{\bp}{{\bf p}}

\newcommand{\br}{{\bf r}}
\newcommand{\bx}{{\bf x}}

\newcommand{\ra}{\rightarrow}

\newcommand{\ve}{\varepsilon}

\newcommand{\wh}[1]{\widehat{#1}}

\newcommand{\om}{{\omega}}

\begin{document}

\centerline{\Large\bf{Electromagnetism and photons in continuous media\footnote{Extended version of talk given at Oberw\"olz meeting, September 7-13, 2008.}}}

\bigskip
\centerline{Finn Ravndal}
\bigskip
\centerline{\it Department of Physics, University of Oslo, Blindern, N-0316 Oslo, Norway.}

\begin{abstract}

\small{Different theoretical and experimental aspects of electromagnetic phenomena in media is reviewed.  The 100 year old Minkowski theory is in agreement with most experiments, but has theoretical problems related to its implied validity in all inertial frames. It is suggested that the similar Abraham theory should be permanently laid to rest since it is not compatible with basic quantum mechanics and is in disagreement with most experiments.  Recently an effective field theory has been proposed which avoids these problems by considering the photon as a quasiparticle like any other excitation in condensed matter physics for which the rest frame of the medium is a preferred frame. It relates many different classical and quantum optical phenomena in a unified description.}

\end{abstract}

\section{Introduction}

In a continuous medium with index of refraction $n$ the velocity of light is $1/n$ in units where it is $c_0 =1$ in vacuum. A monochromatic wave with a frequency $\nu$ thus gets the shorter wavelength $\lambda = 1/n\nu$ when it enters such a medium. This simple fact is the basis of geometrical optics as shown in Fig. 1. 
\begin{figure}[htb]
  \begin{center}
    \epsfig{figure=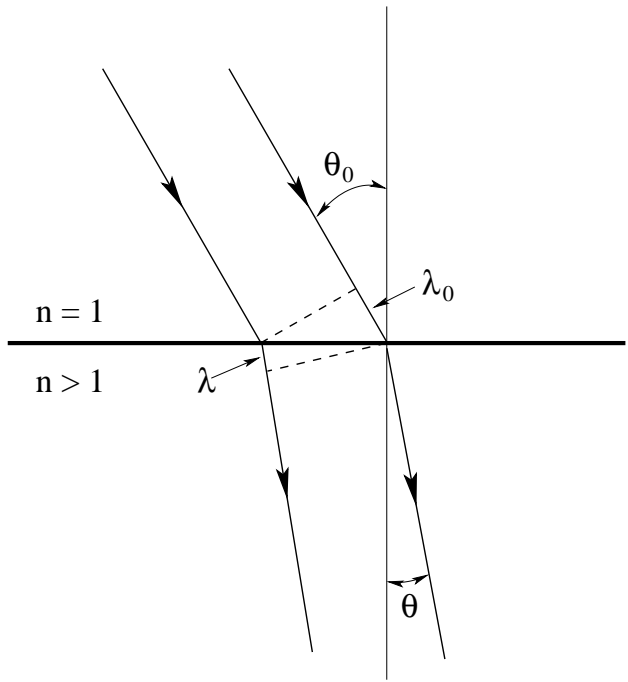,height=90mm}
  \end{center}
  \vspace{-4mm}            
  \caption{\footnotesize When light  enters a denser medium with index of refraction $n$, the vacuum wavelength $\lambda_0$ is shortened to $\lambda = \lambda_0/n$.}
  \label{Fig.1}
\end{figure}
When this electromagnetic wave is quantized, the corresponding photon should have a momentum $p =  h/\lambda$ and an energy $E = h\nu$.   Introducing the wave vector $\bk$ in the direction of the wave, we can then write the momentum vector as $\bp = \hbar\bk$ where the wave number $k = 2\pi/\lambda$. Similarly,  the energy becomes $E = \hbar\om$  where $\om = k/n$.

Let us now consider a couple of  consequences of these simple ideas. In a recent paper by Brevik and Milton\cite{BM} the Casimir force was calculated between two parallel, metallic  plates separated by a distance $L$ and enclosing dielectric matter with a refractive index $n$. After a rather long and detailed calculation they found the resulting force to be a factor $n$ smaller than the standard vacuum force  $F_0 = -\hbar\pi^2/240L^4$. The force results from the zero-point, electromagnetic field energy between the plates which is just $\sum_\bk \hbar\om_\bk = (\hbar/n)\sum_\bk|\bk|$ in the above description. Except for the factor $1/n$, this just the ordinary Casimir energy for vacuum between the plates. We thus have reproduced their result without any calculations.

A related example is black-body radiation in a cavity filled with the same dielectric matter with temperature $T$. Standard statistical mechanics says then that the energy density in the large-volume limit is given by
\be
       u = 2\!\int\!{d^3k\over(2\pi)^3}{\hbar\om_\bk\over e^{\hbar\om_\bk/k_BT} - 1}
\ee
where $k_B$ is the Boltzmann constant. Again using $\om_\bk = k/n$ we find a result which is simply $n^3$ times the vacuum value $u_0 = \pi^2(k_BT)^4/15\hbar^3$. In the book by Landau and Lifshitz  the same result is derived from consideration of correlators of fluctuating currents in the enclosing cavity walls\cite{LL}. At the end of a rather elaborate calculation, they just state without any further comments that the same result can be obtained more directly as done here. It would be interesting and of some importance to verify this experimentally.

These simple ideas thus seem to reproduce some results in a satisfactory way. But can it be part of a consistent theory? What about the photon mass in this picture? In special relativity the squared mass is given by $m^2 = E^2 - p^2$. This gives in our case $(\hbar\om)^2(1 - n^2) < 0$, i.e the photon four-momentum is space-like as for a tachyon.  We will in the following see that this is actually the result emerging from a theory dating back to Minkowski\cite{Minkowski}. Can we live with this result today? Tachyons in ordinary field theories usually signal some instability which we don't expect to find here. And what about gauge invariance? This fundamental symmetry is in vacuum related to having massless photons.

The index of refraction $n$ of a medium gives an effective description valid on large scales where the discrete atoms in the material can be replaced by a continuous medium. On the atomic scale light is moving with the vacuum velocity $c_0 =1$ between interactions with electrons around the atoms. These will scatter the light in such a way that in the forward direction the scattered waves add up to a plane wave. However, it is delayed by  a phase shift of $\pi/2$ relative to the incoming wave as for instance explained by Feynman\cite{RPF}. 
\begin{figure}[htb]
  \begin{center}
    \epsfig{figure=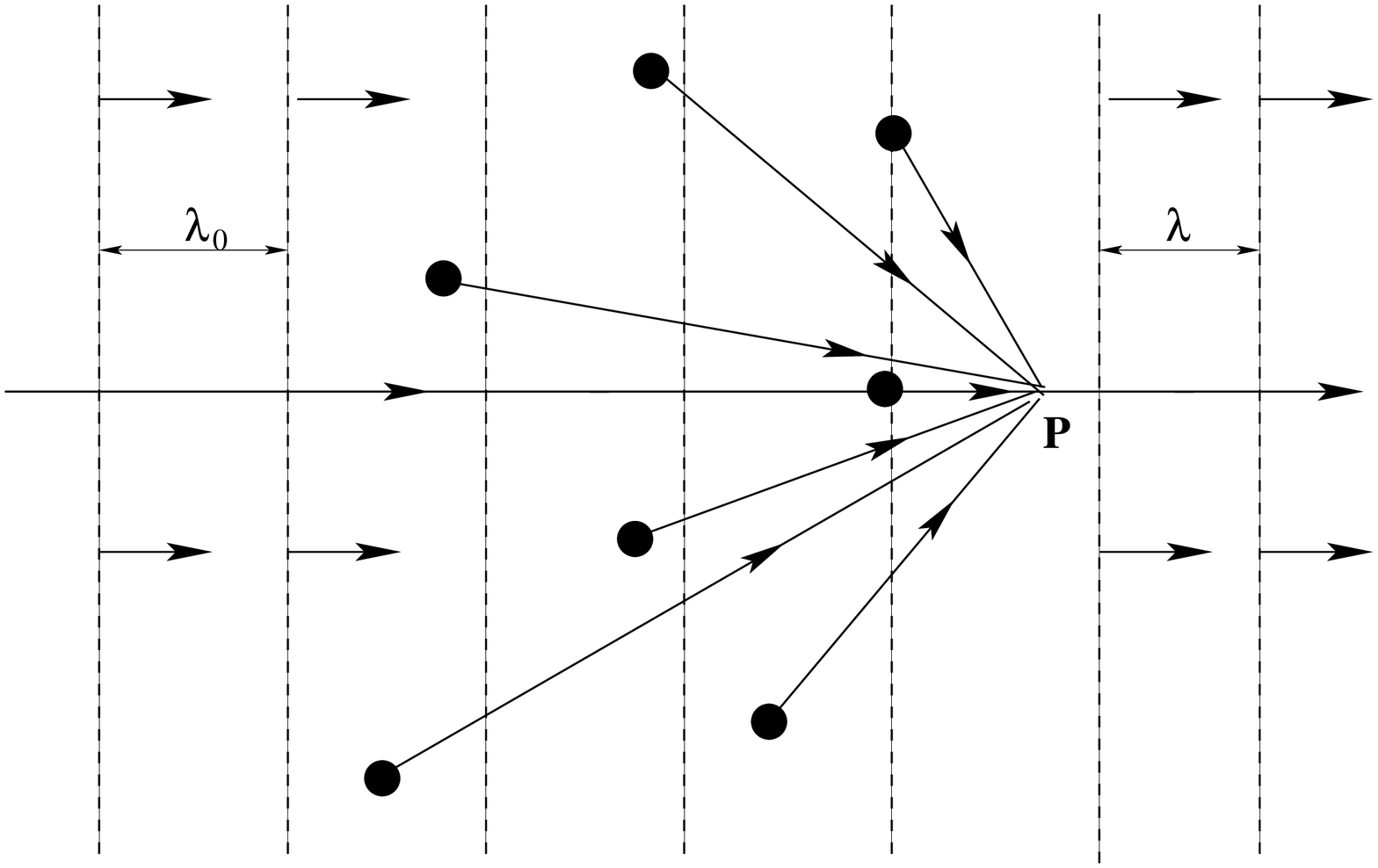,height=70mm}
  \end{center}
  \vspace{-4mm}            
  \caption{\footnotesize The propagation of light in a medium results from an interference between the incoming wave and scattered waves from individual atoms in such a way that its velocity is reduced to $1/n$.}
  \label{Fig.2}
\end{figure}
The interference between these two waves will then effectively slow down the propagating wave.  As a result of these microscopic processes, the resulting wave is therefore a highly complex object. In spite of that, experience shows that we can describe such processes at large scales by local electromagnetic fields obeying the standard Maxwell's equations for continuous media. But these are now effective fields, incorporating complicated physics on very short scales. 

There is no controversy around these macroscopic Maxwell equations. One would then think that a consistent, theoretical description of electromagnetism in continuous media would exist.  But for one hundred years two different theories, one due to Minkowski\cite{Minkowski} and the other to Abraham\cite{Abraham}, have been used and their validity not yet completely settled.  The main difference between them is found in the energy-moment\-um tensors. Standard textbooks\cite{LL}\cite{EM-texts} seem to prefer the Abraham version although it is the Minkowski version which agrees with most experiments\cite{Brevik}\cite{Chiao}. When the theories are quantized, new problems arise.  While the Abraham photon has a three-momentum which seems do be in conflict with basic quantum mechanics, a free photon in the Minkowski theory can have negative energy.

Recently an attempt has been made to clarify this rather confusing situation\cite{FR}. Most of the problems  seem to result from forcing the theory into the standard framework of special relativity which is not present as a physical symmetry of the underlying theory. Instead one can avoid the problems by considering the electromagnetic field in the macroscopic limit as any other excitation in a medium for which the rest frame is a preferred frame. The resulting photons moving with velocity $1/n$ are then quasi-particles on the same footing as quanta in any other field theory with a linear dispersion relation.

Starting with the Maxwell equations in the next section, we review the derivation of the energy and momentum of the electromagnetic field in a continuous medium. The relativistic extensions of this theory proposed by Minkowski  and Abraham are shortly summed up.  In the following section the new, effective theory is presented and then quantized.  As an illustration of the physical consequences of this new description, the following section is devoted to the Cerenkov effect. It can also be understood within the Minkowski theory if a free photon can have negative energy in a frame where the medium is moving.  On the other hand, the Cerenkov effect at the quantum level  is inconsistent with the Abraham formulation. 

In the last section  the Lagrangian for the effective theory is extended with higher order interactions in order to describe non-linear dispersion and the Kerr effects. Thus it relates many different classical and quantum optical phenomena into a unified and consistent theory.

\section{Maxwell theory}

Assuming no charges or currents present in the material,  the electric fields  ${\bf E}, {\bf D}$  and magnetic fields ${\bf B}, {\bf H}$ are in general governed by the  Maxwell equations
\beq
          {\bg\nabla}\times{\bf E} + {\del{\bf B}\over\del t} = 0, \hspace{10mm}      {\bg\nabla}\cdot{\bf B} = 0                     \label{Max-1}
\eeq
and
\beq
          {\bg\nabla}\times{\bf H} - {\del{\bf D}\over\del t} = 0,     \hspace{10mm}    {\bg\nabla}\cdot{\bf D} = 0                 \label{Max-2}
\eeq
The displacement field ${\bf D}$ describes the modification of the electric field ${\bf E}$ by the polarization of the atoms in the material, while ${\bf H}$ describes the similar modification of the magnetic field ${\bf B}$ due to magnetization of the atoms. When the medium can be considered as an isotropic continuum, the relation between these macroscopic fields in the rest frame of the system can be written as ${\bf D} = \ve {\bf E}$ and ${\bf B} = \mu {\bf H}$ as explained in standard text books\cite{EM-texts}. These constitutive relations represent very complex phenomena on a microscopic scale involving a large number of atoms.  The effective  description is therefore only valid on large scales, or equivalently, at sufficiently low energies.  

As a first approximation we will take the electric permittivity $\ve$ and the magnetic permeability $\mu$ to be constants.  In the following we will use units so that for the vacuum $\ve_0 = \mu_0 = 1$. It is then straight-forward to show that the above Maxwell  equations are Lorentz invariant, but only for transformations involving the physical speed of light $1/\sqrt{\ve\mu}$ in the medium. This should be obvious without any explicit derivation since the theory is identical with the one in vacuum except for this difference in light velocity.

Since the second Maxwell equation in (\ref{Max-1}) is satisfied by writing ${\bf B} = {\bg\nabla}\times{\bf A}$ where
${\bf A}$ is the magnetic vector potential, it follows from the first equation that ${\bf E} + \del{\bf A}/\del t$ must be a gradient of a scalar field. One can therefore write
\beq
                  {\bf E} = - {\del{\bf A}\over\del t} - {\bg\nabla}\Phi             \label{E-field}
\eeq
where $\Phi$ is the electric potential. Both the electric and magnetic fields in the medium can therefore be expressed in terms of potentials in the same way as in vacuum. They are invariant under the simultaneous gauge transformations ${\bf A} \ra {\bf A} + {\bg\nabla\chi}$ and $\Phi \ra \Phi - \del\chi/\del t$ where $\chi(\bx,t)$ is an arbitrary, scalar function.

Using now these field expressions together with the constitutive relations in the first of equation (\ref{Max-2}), one obtains the equation of motion
\be
         {\bg\nabla}\times({\bg\nabla}\times{\bf A}) + \ve\mu{\del\over\del t}\Big({\del{\bf A}\over\del t} +{\bg\nabla}\Phi\Big) = 0    
\ee
for the two potentials. Introducing the index of refraction $n=\sqrt{\ve\mu}$, it can be rewritten as
\be
          \Big(n^2{\del^2\over\del t^2} - {\bg\nabla}^2\Big){\bf A} + {\bg\nabla}\Big(n^2\dot{\Phi} + {\bg\nabla}\cdot{\bf A} \Big) = 0
\ee 
Now imposing the gauge condition
\beq
          n^2\dot{\Phi} + {\bg\nabla}\cdot{\bf A} = 0               \label{Lorenz}
\eeq
in the medium, one obtains the standard wave equation
\beq
               \Big(n^2{\del^2\over\del t^2} - {\bg\nabla}^2\Big){\bf A}(\bx,t) = 0     \label{waveq}
\eeq
The electromagnetic propagation velocity is thus $1/n$ as expected.  Needless to say, the gauge condition (\ref{Lorenz})  is equivalent to choosing the covariant Lorenz gauge in vacuum. 

With the assumption of no free charges, the Maxwell equation ${\bg\nabla}\cdot{\bf E} = 0$ gives the relation ${\bg\nabla}\cdot\dot{\bf A} = -\nabla^2\Phi$ with the use of  (\ref{E-field}). Taking the time derivative of the gauge condition (\ref{Lorenz}), we then see that the scalar potential $\Phi(\bx,t)$ satisfies the same wave equation (\ref{waveq}) as the vector potential. Both of these equations of motion follow from the Lagrangian
\beq
             {\cal L} = {1\over 2}\ve {\bf E}^2 - {1\over 2\mu} {\bf B}^2                  \label{L}
\eeq
where the potentials ${\bf A}$ and ${\Phi}$ are the dynamic fields. On this form it is obviously only valid in the rest frame of the medium.

The energy content of the electromagnetic field in a medium is obtained by standard methods\cite{EM-texts}. One takes the scalar products of the first equation in (\ref{Max-1}) with ${\bf H}$ and the first equation in (\ref{Max-2}) with ${\bf E}$. Subtracting the two resulting expressions, the equation
\beq
      {\del{\cal E}\over\del t} + {\bg\nabla}\cdot{\bf N} = 0                                        \label{energy-cons}
\eeq
follows. It represents conservation of energy where 
\beq
          {\cal E} =  {1\over 2} \big({\bf E} \cdot{\bf D} + {\bf B} \cdot{\bf H}\big)          \label{energy}
\eeq
is the standard energy density and ${\bf N} = {\bf E}\times{\bf H}$ is the Poynting vector describing the energy current carried by the field.

Momentum conservation can be similarly obtained by forming the vector products  of the first equation in (\ref{Max-1}) with ${\bf D}$  and the first equation in (\ref{Max-2}) with ${\bf B}$. Combining the two resulting expressions, one then finds  
\be
          ({\bg\nabla}\times{\bf H})\times {\bf B} + ({\bg\nabla}\times{\bf E})\times{\bf D} =  {\del\over\del t} ({\bf D}\times{\bf B})
\ee
This can be written on a more compact form using the triple vector product formula ${\bf A}\wedge({\bf B}\wedge{\bf C}) = ({\bf A}\cdot{\bf C}){\bf B} -  ({\bf A}\cdot{\bf B}){\bf C}$. It results in
\beq
        {\del{\bf G}\over\del t} + {\bg\nabla}\cdot{\bf T} = 0    \label{mom-cons}
\eeq
where ${\bf G} = {\bf D}\times{\bf B}$ and
\beq
                T_{ij} =  - (E_i D_j + B_i H_j) + {1\over 2}\delta_{ij}({\bf E} \cdot{\bf D} + {\bf B} \cdot{\bf H})     \label{max-stress}
\eeq
is the Maxwell stress tensor. Using the constitutive equations, it is seen to be symmetric in the rest frame of the medium. It is thus natural to consider the vector ${\bf G}$ to represent the momentum density of the field.

\section{Minkowski and Abraham formulations}

There seems to be no disagreement around the presentation  given in the previous section. The difficulties start when one attempts to embed this non-covariant formulation into a four-dimensional framework based on the special theory of relativity. One could then discuss electromagnetic phenomena in a general, inertial frame where the medium could have any velocity below the velocity of light in vacuum. This was first done by Minkowski at the same time as his successfull covariant formulation of the Maxwell theory in vacuum was completed\cite{Minkowski}. 

In the rest frame of the medium the space and time coordinates of an event can be combined into a four-dimensional vector
$x^\mu = (t, \bx)$.  The covariant gradient operator is then $\del_\mu = (\del/\del t, {\bg\nabla})$. We will raise and lower Greek indices with the standard Lorentz metric $\eta_{\mu\nu}$, taken here to have negative signature. Combining the two potentials $\Phi$ and ${\bf A}$ into the four-dimensional vector potential $A^\mu = (\Phi, {\bf A})$, the antisymmetric field tensor $F_{\mu\nu} = \del_\mu A_\nu - \del_\nu A_\mu$ is then seen to have the components
\beq
           F^{\mu\nu} = \left( \begin{array}{c | c} 0 & - {\bf E} \\ \hline  {\bf E} & - B_{ij}\end{array}\right)          \label{F}
\eeq
in the same frame where $B_{ij} = \ve_{ijk} B_k$. The first two Maxwell equations (\ref{Max-1}) can then be written as
\beq
            \del_\lambda F_{\mu\nu} +  \del_\nu F_{\lambda\mu} +  \del_\mu F_{\nu\lambda} = 0         \label{max-1}
\eeq
Thus this part of the Maxwell theory in a medium is the same as in vacuum. 

The problems arise with the remaining fields ${\bf D}$  and ${\bf H}$. In analogy with the tensor (\ref{F}) they can be combined into a new, antisymmetric tensor
\beq
           H^{\mu\nu} = \left( \begin{array}{c | c} 0 & - {\bf D} \\ \hline {\bf D} & - H_{ij}\end{array}\right)          \label{H}
\eeq 
with $H_{ij} = \ve_{ijk} H_k$. The two last Maxwell equations (\ref{Max-2}) can then be simply reduced to $\del_\nu H^{\mu\nu} = 0$.  It also makes it possible to write the Lagrangian (\ref{L}) on the compact form ${\cal L} = - F_{\mu\nu}H^{\mu\nu}/4$ when we make use of the constitutive equations in the rest frame. 

Despite the covariant form of the Lagrangian, it does not represent a Lorentz-invariant theory in the usual sense. This is so because the phenomenological tensor $H_{\mu\nu}$ must be expressed in terms of the more fundamental tensor $F_{\mu\nu}$  in a frame where the medium is in motion so to generalize the constitutive equations ${\bf D} = \ve {\bf E}$ and ${\bf B} = \mu {\bf H}$ valid only in the rest frame of the medium. Such a relation can always be found, but will obviously involve the velocity of the medium\cite{Minkowski}\cite{JW-1}. In a general frame this velocity will then enter the Lagrangian explicitly and thus signal the lack of physical invariance under vacuum Lorentz transformations as already mentioned. The true invariance of the Maxwell theory in a medium is represented by Lorentz transformations involving the reduced speed of light $1/n$.

But as long as we restrict ourselves to the rest frame of the medium, there are so far no problems. The energy and momentum content of the field derived in the previous section, can then be combined into the four-dimensional energy-momentum tensor
\beq
           T^{\mu\nu}_M =  \left( \begin{array}{c | c} {\cal E} & {\bf N} \\ \hline  {\bf G} &T_{ij}\end{array}\right)        \label{T_Mmat}
\eeq
valid in this frame. Minkowski wrote it as
\beq
       T^{\mu\nu}_M =  F^\mu_{\,\,\,\alpha}H^{\alpha\nu} + {1\over 4}\eta^{\mu\nu} F_{\alpha\beta}H^{\alpha\beta}      \label{T_M}
\eeq
with the intention of making use of it in any inertial frame. A direct derivation can be found in the book by M\"oller\cite{Moller}. The two  conservation laws can now be expressed on the more compact  form $\del_\nu T^{\mu\nu}_M = 0$.  This energy-momentum tensor is seen in general not to be symmetric, i.e. the total angular momentum  of the field is not conserved.  Only in the limit $n\ra 1$ where it becomes the electromagnetic energy-momentum tensor of the vacuum, do we recover this desired property.

In order to remedy this lack of symmetry,  Abraham proposed the following year that only the symmetric part of the Minkowski energy-momentum should be used\cite{Abraham}.  More formally, this was done by splitting the Minkowski tensor (\ref{T_Mmat}) into two terms,
\be
          T^{\mu\nu}_M =   T^{\mu\nu}_A +  (n^2 - 1)\left( \begin{array}{c | c} 0 & 0\\ \hline  {\bf N} & 0 \end{array}\right)   
\ee
where the first part
\beq
            T^{\mu\nu}_A =  \left( \begin{array}{c | c} {\cal E} & {\bf N} \\ \hline  {\bf N} &T_{ij}\end{array}\right)      \label{T_A}
\eeq
is the Abraham energy-momentum tensor. It is symmetric by construction. From the above conservation law for the Minkowski tensor on the second index, it follows that the Abraham tensor is not generally conserved. Instead, it  is seen to satisfy $\del_\nu T^{\mu\nu}_A +  K^\mu = 0$ where $K^\mu$ is a force density. In the same rest frame as above we have $K^\mu = (0,{\bf K})$ where
\beq
         {\bf K} = (n^2 - 1){\del\over\del t}({\bf E}\times{\bf H})           \label{A-force}
\eeq
is the Abraham force. The time component of this new conservation law ensures energy conservation on the standard form (\ref{energy-cons}). While the spatial components again ensure momentum conservation, the momentum density of the field is now seen to be ${\bf E}\times{\bf H}$ and therefore $n^2$ smaller than the above Minkowski density ${\bf D}\times{\bf B}$.  This will result in a correspondingly smaller radiation pressure and seems to be ruled out by most experiments\cite{pressure}.  For this reason it has been suggested that in some way the motion of the microscopic matter should be included in this formalism to give an effective momentum density equal to the Minkowski theory\cite{Blount}.  But according to Garrison and Chiao\cite{Chiao}  the Abraham formalism still seems to be needed to explain a few experiments where the systems under investigation undergo acceleration.  Different theoretical  approaches to these problems have recently been reviewed by Obukhov\cite{Obukhov}.

\section{The effective field theory}

Both the Minkowski and the Abraham formulations are based on being valid in any inertial frame related by ordinary vacuum Lorentz transformations. During the last 100 years the controversy around these two theories has not been settled. As a way out of this impasse, it has recently been proposed to consider these electromagnetic phenomena in the same way as  other excitations described by field theory in condensed-matter physics\cite{FR}. Since the medium itself has properties which are not invariant under Lorentz transformations, it is natural and most common to limit such descriptions to the rest frame of the medium where the field theories are defined.

In this frame light moves with the velocity $1/n$. The corresponding light cone is $|\bx | = \pm t/n$.  As in vacuum, it is desirable to write this on an infinitesemal level as $ds^2 = 0$ with a line element  on the form $ds^2 = \eta_{\mu\nu}dx^\mu dx^\nu$. If we now choose $\eta_{\mu\nu}$ to be the Minkowski vacuum metric, the contravariant coordinates in this frame must be $x^\mu = (t/n, \bx)$. The corresponding covariant derivative is obviously then $\del_\mu = (n\del/\del t, {\bg\nabla})$. In a quantum theory this should correspond to the four-momentum $p^\mu = (nE,\bp)$ for a particle with energy $E$ and three-momentum $\bp$. The d'Alembertian $\del^\mu\del_\mu = (n^2\del_t^2 - \nabla^2)$ is invariant under Lorentz transformation corresponding to the light speed $1/n$. It is seen to equal the wave operator we found for the Maxwell theory in a medium.

This theory can now be given a simple covariant formulation. We introduce a four-vector electromagnetic  potential $A^\mu = (n\Phi, {\bf A})$ so that the electric and magnetic field vectors are again given by the antisymmetric tensor $F_{\mu\nu} = \del_\mu A_\nu - \del_\nu A_\mu$. It has now the components
\beq
           F^{\mu\nu} = \left( \begin{array}{c | c} 0 & -n {\bf E} \\ \hline  {n\bf E} & - B_{ij}\end{array}\right)          \label{new-F}
\eeq
instead of (\ref{F}) for the Minkowski formulation.  The rest-frame Lagrangian (\ref{L}) takes then the standard form $\mu{\cal L} = - (1/4)F_{\mu\nu}^2$.  The first set of field equations (\ref{max-1}) obviously remains unchanged while the second Maxwell equations (\ref{Max-2}) are replaced by $\del_\mu F^{\mu\nu} = 0$ when we make use of the con\-stitutive equations.  One thus obtains the wave equation $\del^2 A^\nu - \del^\nu(\del\cdot A) = 0$.  In the Lorenz gauge  defined by $\del_\mu A^\mu = 0$, it gives the previous wave equation (\ref{waveq}). Notice that this covariant gauge condition becomes ({\ref{Lorenz}) when written out in terms of components.

From the above invariant Lagrangian the energy-momentum tensor can now be derived as in vacuum, giving
\beq
                \mu T^{\mu\nu} =  F^\mu_{\,\,\,\alpha}F^{\alpha\nu} + {1\over 4}\eta^{\mu\nu} F_{\alpha\beta}F^{\alpha\beta}      \label{T_eff}
\eeq
with components
\beq
           T^{\mu\nu} =  \left( \begin{array}{c | c} {\cal E} & n {\bf N} \\ \hline n {\bf N} &T_{ij}\end{array}\right)        \label{T_effmat}
\eeq
It is obviously symmetric, traceless and conserved on both indices, i.e. $\del_\mu T^{\mu\nu} = \del_\mu T^{\nu\mu} = 0$. In the time direction this gives energy conservation on the form (\ref{energy-cons}) while in the space directions it gives momentum conservation as in (\ref{mom-cons}). The momentum density of the field ${\bf G} = {\bf D}\times{\bf B}$ is therefore the same as in standard Maxwell theory and for the Minkowski description restricted to the rest frame.  

In this frame we have the Lagrangian density (\ref{L}) and the theory can be quantized by standard methods. With no free charges, we can take the scalar potential $\Phi = 0$ and use the Coulomb gauge ${\bg\nabla}\cdot{\bf A} = 0$. There are then only two transverse field degrees of freedom governed by Lagrangian
\beq
             L =  \int\!d^3x\left[{1\over 2} \ve{\dot{\bf A}}^2 - {1\over 2\mu}({\bg\nabla}\times{\bf A})^2 \right]
\eeq
With the system in a volume $V$ with periodic boundary conditions, we can expand the vector potential in plane waves as
\beq
             {\bf A}(\bx,t) = \sqrt{1\over V}\sum_\bk  {\bf A}_\bk(t) e^{i\bk\cdot\bx}                     \label{A_class}
\eeq
where each Fourier mode with amplitude  ${\bf A}_\bk(t)$ is characterized by a discrete wave vector $\bk$. In terms of these complex amplitudes satisfying ${\bf A}_{\bk}^* = {\bf A}_{-\bk}$,  the Lagrangian  becomes
\be
        L = {1\over 2}\ve \sum_\bk\left(\dot{\bf A}_\bk \dot{\bf A}_{\bk}^* - \om_\bk^2{\bf A}_\bk\cdot{\bf A}_\bk^*\right)
\ee
Each term is seen to describe a harmonic oscillator with frequency $\om_\bk = |\bk|/n$. Introducing creation and annihilation operators for photons with definite polarizations $\lambda$, the quantized Hamiltonian takes the standard form
\beq
         H = \sum_{\bk\lambda}\hbar\om_\bk \left({a}_{\bk\lambda}^\dagger{a}_{\bk\lambda} + {1\over 2}\right)      \label{EM_Ham}
\eeq
where the last term gives the zero-point energy. A single photon with the wave vector $\bk$ thus has the energy $E = \hbar\om_\bk$. This will also be the photon energy in the Minkowski and Abraham theories as long as they are restricted to the rest  frame of the medium.

With the above classical momentum density, we can now find the operator for the total momentum of the quantized field from
\beq
            {\bf P} = \int\! d^3x\, {\bf D}\times{\bf B}
\eeq
It simplifies to 
\beq
      {\bf P} = \sum_{\bk}\hbar\bk (a_{\bk +}^\dagger {a}_{\bk +} + a_{\bk -}^\dagger {a}_{\bk -})           \label{Mom-op}
\eeq
when we make use of the same plane-wave expansion and write out explicitly the contributions from the two polarization directions. A photon with wave vector $\bk$ has therefore the momentum $\bp = \hbar\bk$.  Since it has  the energy $E = \hbar\om_\bk$, its squared four-momentum is $(nE)^2 - p^2 = 0$. Thus it can be said to be massless  in a medium when described by the effective theory. This is in contrast to the Minkowski theory where the mass-squared of the same photon would be $E^2 - p^2 = (\hbar\om)^2(1-n^2)  < 0$, while it is  $(\hbar\om)^2(1-1/n^2)  > 0$ in the Abraham formulation.

The total angular momentum of the field is given by the classical expression
\beq
                 {\bf J} = \int\! d^3x\, \br\times ({\bf D}\times{\bf B})
\eeq
Separating out the orbital part, the intrinsic spin part can be quantized and becomes
\beq
                 {\bf S} =  \sum_{\bk}\hbar \wh{\bk}(a_{\bk +}^\dagger {a}_{\bk +} - a_{\bk -}^\dagger {a}_{\bk -})
\eeq
where $\wh{\bk}$ is a unit vector along the wave vector $\bk$. Needless to say, this is exactly the same result as in vacuum. The photon in a medium  thus has spin $S = 1$ with only two helicities $\lambda = \pm$ required for a massless vector particle. 

In the Minkowski formulation the photon has a non-zero mass and one should therefore {\it a priori} expect the spin to have a third direction. This is even more true for the Abraham formulation, but here the magnitude of the photon spin is reduced to $S = 1/n^2$. It was therefore suggested by Brevik in his review paper\cite{Brevik}) that a measurement of the photon spin would offer a clear method to differentiate between these two theories. Some years later such an experiment was performed\cite{KW}  giving a value very close to $S=1$. Even if this measurement was  not made on free photons as above, but on photons in a wave guide filled with a dielectric liquid, the result should be the same. Again the validity of the Abraham theory seems to be ruled out.

\section{Cerenkov radiation of photons}

While the effective theory is only valid in the medium rest frame, the Minkowski formulation is by construction valid in any inertial frame related to the rest frame by a vacuum Lorentz transformation. The theory can then in principle be quantized in such an arbitrary frame where the medium is in motion. This was first done by Jauch and Watson\cite{JW-1}. As expected, it is much more cumbersome than the above rest-frame quantization and with new problems. This should not come as a surprise since these  vacuum Lorentz transformations do not represent a physical symmetry. In particular there are difficulties in the treatment of the longitudinal components of the radiation field. A later attempt by Brevik and Lautrup to clarify the situation, did not lead to a definite conclusion\cite{BL} .

A simple example of such a problem is to consider a photon with the four-momentum $p^\mu = (\hbar\om_\bk,\hbar\bk)$ moving with velocity $1/n$ along the $x$-axis in the rest frame. As pointed out above, this four-momentum is space-like. Now going to a new inertial frame by a vacuum Lorentz transformation moving along the $x$-axis with a velocity $v > 1/n$, it will be observed to have negative energy\cite{BH}. What this means physically, is not clear. One cannot simply say it has a negative frequency. It must in some way be the matter which zooms by in this frame, which imparts upon the photon  this negative energy.  In practice we meet the same problem when Cerenkov radiation is explained within the Minkowski theory.

When a charged particle with a speed $v > 1/n$ passes through a medium with index of refraction $n$, electromagnetic radiation is emitted. This Cerenkov effect is similar to a sonic boom when an object goes through air with a speed larger than the speed of 
\begin{figure}[htb]
  \begin{center}
    \epsfig{figure=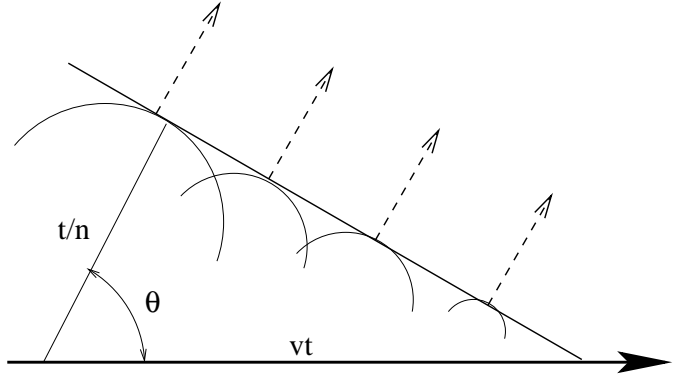,height=50mm}
  \end{center}
  \vspace{-4mm}            
  \caption{\footnotesize Cerenkov radiation from particle with velocity $v$ during a time $t$ in a medium where velocity of light is $1/n$.}
  \label{Fig.1}
\end{figure}
sound. The radiation is emitted in a cone with opening angle given by $\cos\theta = 1/nv$ as shown in Fig.3. As first demonstrated by Frank and Tamm, it is a classical effect following directly from the previous Maxwell equations in the rest frame of the medium\cite{EM-texts}. 

At the microscopic level it corresponds to the incoming particle emitting a photon in a direction $\theta$ away from the incoming direction and continuing in a slightly different direction with smaller energy as shown in Fig.4. 
If we denote the energy and momentum of the incoming particle by $E$ and  $p$ and similarly primed quantities for the outgoing particle, then energy conservation implies $E = E' + \hbar\om$. The photon frequency $\om$ is related to its wave number by $\om = k/n$. Using now the photon momentum $\hbar k$ from the previous section, one has momentum conservation $p = p_x' + \hbar k\cos\theta$ along the incoming $x$-direction. In the normal $y$-direction, it similarly follows that $p_y' + \hbar k\sin\theta = 0$. Squaring these two eqations and adding, it follows that
\beq
               p'^2 = p^2 + (\hbar k)^2 - 2\hbar kp\cos\theta
\eeq
\begin{figure}[htb]
  \begin{center}
    \epsfig{figure=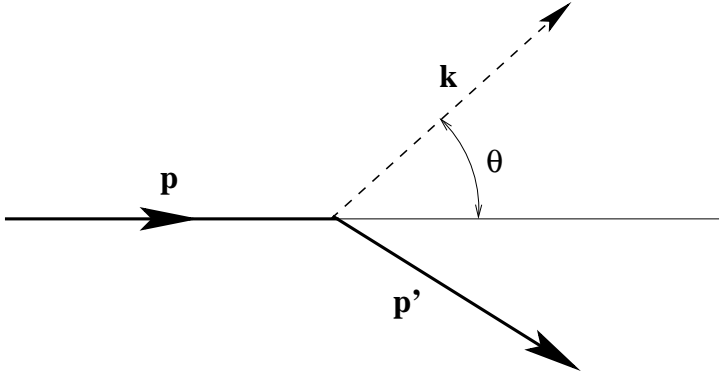,height=50mm}
  \end{center}
  \vspace{-4mm}            
  \caption{\footnotesize Cerenkov radiation of a photon with wave vector $\bk$ from a charged particle with momentum $\bp$.}
  \label{Fig.2}
\end{figure}
Combining this with the squared conservation equation for energy which takes the form $p'^2 = p^2 - 2\hbar k E/n + (\hbar k/n)^2$, the deflection angle is seen to be determined by
\beq
            \cos\theta = {1\over nv} + {\hbar k (n^2-1)\over 2pn^2}
\eeq
where $v = p/E$ is the velocity of the incoming particle. When the particle is relativistic and we consider the emission of visible light, the last, quantum term can be neglected and the photon angle is given by the classical expression. The quantum mechanical transition rate for the process was calculated by Jauch and Watson in the same frame using the Minkowski formulation\cite{JW-2}. They obtained a radiation rate in agreement with the Frank-Tamm result. This is to be expected from the correspondence principle. 

However, within the Minkowski formalism one can in principle consider the process in any other inertial frame where the theory should be just as valid. For this reason Jauch and Watson also used the special frame where the incoming particle is at rest. From the kinematics in this frame it then follows that this particle can decay into a new particle with a certain three-momentum plus a photon with the opposite momentum. Since the masses of the initial and final particles are assumed to be the same, energy conservation then gives that the photon must have negative energy in this frame. It is therefore a photon with properties very different from all other photons in physics. Its strange properties must result is some way from the matter of the medium zooming by in this frame. Although this result seems to be mathematically correct, one must be allowed to ask about its physical validity.

For the Abraham description this process is catastrophic. Going through the same steps as above, but now with the photon momentum $\hbar k/n^2$, it follows immediately that the classical term for the deflection angle gives $\cos\theta = n/v > 1$ for physical velocities. Thus there can be no Cerenkov effect at the quantum level in this case as also noted a long time ago by Brevik and Lautrup\cite{BL}. This should come as no surprise since the momentum of a photon with wavelength $\lambda$ is no longer given by the fundamental de Broglie expression $h/\lambda$ in this formulation. 

\section{Higher order interactions}

So far we have only considered the free theory described by the Lagrangian (\ref{L})  and assuming the phenomenological parameters $\ve$ and $\mu$ to be constants. It is therefore only valid on very large scales, i.e. at energies so low that no microscopic degrees of freedom are excited. For a physical medium made out of atoms this corresponds to energies much less than a few eV. At higher energies, these effects will start to manifest themselves and must be included some way. In particular we need to incorporate non-linear dispersion in order for the theory to be realistic. And it must be done in such a way that it allows a consistent treatment at the quantum level.

The free theory was formulated along the same lines as for other excitations in condensed matter physics. For many years it has been well known in this field how to incorporate microscopic effects in a macroscopic description by extending the free theory in the rest frame by including higher-order operators in the Lagrangian. The coupling constants of these new terms are determined by the microscopic physics. They must be determined from an underlying, more fundamental theory or from experiments. The resulting Lagrangian describes then an interacting, effective theory. Although it is in general said to be non-renormalizable, finite quantum corrections can be derived from it as long as one restricts oneself to phenomena below a characteristic energy.  Such effective field theories have during the last 10-20 years become of great use also in high energy physics\cite{Cliff}. The first well-known theory of this kind was found by Euler and Heisenberg already in 1936 to describe classical electromagnetic effects in strong fields, induced by virtual electron-positron  pairs in the vacuum\cite{EH}. It was first quite recently that it became clear that it could also be used as an effective, quantum field theory\cite{KR}.  Now a similar, effective theory  for electromagnetic phenomena in media has been proposed\cite{FR}.

In order to be gauge invariant, higher-order couplings in the Lagrangian can only involve the fields  ${\bf E}$ and  ${\bf B}$ and derivatives of them. For the sake of counting, we can use quantum units with  $\hbar =1$ so that these fields have dimension +2 and every derivative corresponds to an increase in dimension by +1.  To be invariant under time-reversal and ordinary rotations, such new couplings must involve at least two spacetime derivatives. For example, one possibility could be the term ${\bf E}\cdot\del^2{\bf E}$. It has dimension 6. But the lowest order equation of motion is just $\del^2{\bf E} = 0$ and this term can therefore not contribute.  Possible new terms of dimension 8 would be $({\bf E}\cdot{\bf E})^2$,  $({\bf B }\cdot{\bf B})^2$,  ${\bf E}^2{\bf B}^2$ and $({\bf E}\cdot{\bf B})^2$. All such terms describe anharmonic interactions involving four fields. 

The simplest form of non-linear dispersion follows from dimension-6 interactions when we restrict ourselves to a theory with only rotational invariance. One example of a possible interaction is then $\nabla_i{\bf E}\cdot\nabla_i{\bf E}$. It is equivalent to ${\bf E}\cdot\nabla^2{\bf E}$ by a partial integration in the action integral where it appears. The similar term  $\del_t{\bf E}\cdot\del_t{\bf E}$ involving two time derivatives is for the same reason equivalent to ${\bf E}\cdot\nabla^2{\bf E}$ when we use the equation of motion. An interaction like ${\bf E}\cdot\nabla^2{\bf B}$ is ruled out by parity invariance.

Of most interest are dielectric media for which we can set the permeability $\mu = 1$. In such materials magnetic effects are negligible and it is therefore reasonable to assume that all the terms involving the magnetic field, are absent.  The effective Lagrangian then becomes
\beq
           {\cal L} = {1\over 2}\Big(n^2{\bf E}^2 - {\bf B}^2\Big)  +  {d_1\over M^2} {\bf E}\cdot\nabla^2{\bf E}  
                       + {d_3\over M^4} ({\bf\nabla^2 E})^2  + {a_1\over M^4} ({\bf E }\cdot{\bf E})^2                                 \label{L_diel}
\eeq
when we restrict ourselves to operators with dimension 8 or less. $M$ is a characteristic energy below which the theory should be valid. In addition, it contains only three independent dimensionless parameters $d_1$, $d_3$ and $a_1$. For each material they can therefore be determined by three different measurements when the value of $M$ is known. The Lagrangian should then be able to predict the outcome of other experiments  without  any more parameter fitting. 

The effect of the first new term proportional with $d_1$ is simplest to analyze since it is quadratic in the field. In the quantum treatment it will give a perturbation $\Delta E$ to the energy of a photon with momentum $\hbar k$. It is simple to calculate and the result is found to be $\Delta E = - d_1k^3/2M^2n^3$. The resulting total energy $E' = E + \Delta E$ can now be written as $E' = \hbar k/n(\om)$ where the modified index of refraction is
\beq
             n(\om) = n\Big( 1 - {d_1\om^2\over 2M^2}\Big)
\eeq
where $n = \sqrt{\ve}$ as before. Thus it gives the Cauchy parametrization of non-linear dispersion valid for the longest wavelengths of light\cite{AN} when $d_1$ is negative. Comparing with measured values, we find that $M = 5-10$ eV for typical materials if we set the unknown parameter $d_1 = - 1$. The operator  proportional to $d_3$ (\ref{L_diel}) will obviously give a $\om^4$ correction to this dispersion law. Similarly we can show that the operator $({\bf E }\cdot{\bf E})^2$ describes the AC and DC Kerr effects\cite{AN}. The mass parameter $M$ is again found to be in the same range as above if we choose $a_1 =1$\cite{FR}. One can therefore instead take $M$ to have the same value for all materials and let the dimensionless parameters $d_1, d_3$ and $a_1$ vary from material to material.

\section{Conclusion}

The Abraham description of electromagnetism in media is inconsistent with both basic theoretical ideas and experimental results. After having been considered now for 100 years, it is time for it to be laid permanently to rest. While the energy and momentum content resulting from the Minkowski theory in the rest frame of the medium avoid these problems, it still has difficulties with the requirement of being valid in any inertial frame. 

Considering instead these fields like other excitations in condensed matter physics and defined by an effective theory in the medium rest frame,  these problems are avoided. Except for the reduced velocity of light, it is very similar to the corresponding theory in vacuum.  This new theory thus  becomes equivalent to electromagnetism in the ether before 1905.  The Maxwell equations were then considered to be valid only in the rest frame of the ether. Einstein's special theory of relativity showed that there is no need for a physical ether and Maxwell equations became valid in all inertial frames. Today we would rather say that there is an ether, but that it is invariant under Lorentz transformations. In contrast, a physical medium is not invariant under these transformations and that makes the whole difference.

We want to thank I. Brevik and Y. Galperin for several useful discussions.

\end{document}